\documentstyle[12pt]{article}
\def \beq{\begin{equation}}
\def \eeq{\end{equation}}
\def \beqa{\begin{eqnarray}}
\def \eeqa{\end{eqnarray}}
\def \tht{\theta}
\def \mon{\tilde m(\star x)}
\def \vor{\tilde m_{\mu}(\star x)}
\def \curr{m_{\mu}(x)}

\begin{document}
\baselineskip=24pt
\begin{center}
\bf{A lattice model in three dimensions with a $\theta$ term
} \\
\vspace{1cm}

\rm{Srinath Cheluvaraja} \\
Institute of Mathematical Sciences \\
C.I.T. Campus, Tharamani \\
Chennai 600 113, India

\end{center}
\vspace{1cm}
We study a three-dimensional
abelian lattice model in which the analogue of a
theta term can
be defined. This term is defined by introducing a
neutral scalar field and its effect is to 
couple magnetic monopoles
to the scalar field and vortices to the gauge field.
An interesting
feature of this model is the presence of an exact duality symmetry
that acts on a three parameter space.
It is shown that
this model has an
interesting phase structure for non-zero values of
$\theta$. In addition to the usual confinement and  vortex phases
there are phases in which loops with composite charges condense.
The presence of novel point like excitations also alters the physical
properties of the system.


\vspace{5cm}
IMSC/99/05/19

\newpage
\begin{section}{Introduction}
Topology has played an important role in statistical mechanics and
field theory. The topology of the dynamical variables leads to
novel excitations that can have a profound effect in determining the
physical properties of the system. A well known example in statistical
mechanics is the vortex excitation \cite{vort}
in  the two-dimensional planar model.
These vortex excitations, which exist
because of the angular nature of the spin variables,
drive a phase transition which is very different from other
phase transitions in statistical mechanics. In gauge theories 
the analog of the vortex excitation is the
magnetic monopole \cite{hopol}. 
The magnetic monopole exists as a consequence of the
topology of the gauge group. Mult-monopole states are also present as
collective excitations.
These multi-monopole excitations can form a plasma phase
in which electric charges are confined. A classic example of this
phenomenon is compact QED in three Euclidean 
dimensions \cite{poly}. 
Apart from the intrinsic topology of the dynamical variables, we can also
consider terms which have a direct topological significance. One such term is the
well known $\theta$ term in non-abelian gauge theories. 
For the $SU(2)$ non-abelian gauge theory this term is 
defined as
\beq
\label{inst}
\frac{\theta}{32\pi^2}\int d^{4}x\ tr\ F_{\mu \nu} \tilde F_{\mu \nu}
\quad .
\eeq
Although this term is a total derivative it can have a non-trivial effect
whenever long range fields are
present. 
In the space of finite action configurations, the above
term
is precisely the winding number of mappings from
$S_{3}$ to $S_{3}$. Gauge field configurations which give a non-zero 
value for the expression in Eq.~\ref{inst} can affect some of the
physical properties of the system \cite{instant}.
An important effect of the term in Eq.~\ref{inst} is that it can
convert a magnetic monopole into a particle with an additional electric
charge - a dyon \cite{witt}.
It was conjectured in \cite{conj} that non-abelian gauge theories could
have new phases of oblique confinement as a result of the interactions
between these dyons.
The analog of the term in Eq.~\ref{inst} in an
abelian gauge theory is
\beq
\label{fourabe}
\int d^{4}x F_{\mu \nu} \tilde F_{\mu \nu}
\quad .
\eeq
Unlike as in the non-abelian theory, this term does not have a topological
significance as the winding number of any mapping. However, this term can 
have a non-trivial physical effect in the presence of magnetic monopoles.
The effect of the $\theta$ term in Eq.~\ref{fourabe} was studied on the lattice
in
\cite{cardy1}. It was shown that the $\theta$ term drastically alters
the phase structure of the theory and a rich phase structure was uncovered
as a function of $\theta$. The oblique confinement phases conjectured
in \cite{conj} were also elucidated. An exact duality symmetry ( which
is the action of the group $SL(2,Z)$ ) was demonstrated 
to hold in this model \cite{cardy2}
and the action of this symmetry was used to predict the entire phase
structure of the model.
Both the $\theta$ terms discussed so far require four Euclidean dimensions.
In \cite{cardy1,cardy2} some two dimensional spin models were also
considered in which a $\theta$ term could be defined.
We would like to consider such a  term in three Euclidean dimensions and
study its effects.
An inspection of the properties of the
$\epsilon$ tensor shows that defining gauge invariant terms using
only pure gauge fields  is not possible. An exception is the Cherns-Simons
term about which we will comment later.
The $\theta$ term that we will consider in three dimensions is the
lattice analogue of the following term
\beq
\label{3tht}
i\theta \int d^3{x}\epsilon_{\mu \nu \lambda}F_{\mu \nu}\partial_{\lambda}\phi
\quad .
\eeq
It is clear that under the parity transformation in three-dimensions
\beqa
\nonumber
x\rightarrow x \\ \nonumber
y\rightarrow -y \\ \nonumber
t\rightarrow t
\quad ,
\eeqa
the $\theta$ term changes sign i.e $\theta\rightarrow -\theta$.
Therefore, the $\theta$ term violates parity unless the free energy is an
even function of $\theta$.
The field $\phi$ is a neutral field and hence the above term is gauge
invariant. A simple integration by parts gives the following terms
\beq
\partial_{\lambda}(\epsilon_{\mu \nu \lambda}F_{\mu \nu}(x)\phi(x))-
(\epsilon_{\mu \nu \lambda}\partial_{\lambda}F_{\mu \nu}(x))\phi(x)
\quad .
\eeq
The dual of the field strength is defined as
\beq
\tilde F_{\lambda}(x)=\frac{1}{2}\epsilon_{\lambda \mu \nu}F_{\mu \nu}
\eeq
The second term is seen to be $2\partial_{\lambda}\tilde F_{\lambda}(x)\phi(x)$ 
and
this term can be non-zero in the presence of magnetic monopoles.
In the presence of magnetic monopoles 
\beq
\partial_{\lambda}\tilde F_{\lambda}=\tilde m(x)
\eeq
where $\tilde m(x)$ is the magnetic monopole density at $x$.The second term
leads to the coupling
\beq
2\tht i \tilde m(x) \phi(x)
\quad .
\eeq
Another non-trivial contribution to Eq.~\ref{3tht} can come from a 
vortex line.
This is seen by doing the integration by parts in Eq.~\ref{3tht} differently
as
\beq
2\epsilon_{\mu \nu \lambda}\partial_{\mu}(A_{\nu}\partial_{\lambda}\phi)-
2\epsilon_{\mu \nu \lambda}A_{\nu}\partial_{\mu}\partial_{\lambda}\phi
\eeq
The second term can be written as
\beq
-2\theta i\int m_{\mu}(x)A_{\mu}(x)
\eeq
The quantity 
\beq
m_{\mu}(x)=\epsilon_{\mu \lambda \nu}\partial_{\lambda} \partial_{\nu}\phi
\eeq
is non-zero around a vortex line.
Hence, the term in Eq.~\ref{3tht} is non-zero in the presence of magnetic
monopoles and vortex lines and introduces new couplings between the
topological ( $\vor \ \  and\ \ \mon$ )
and the spin wave ($A_{\mu}\ \  and\  \phi$) 
degrees of freedom.
In the absence of these topological excitations,
the term in Eq.~\ref{3tht} will have no physical effect.
Another way of motivating the term in Eq.~\ref{3tht} is by dimensionally
reducing the four-dimensional term $\theta$ term in Eq.~\ref{fourabe}.
At high temperatures, the leading order contribution from a term like
Eq.~\ref{fourabe} is given by Eq.~\ref{3tht}.
In this paper, we will present an analysis of the $\theta$ term in three
Euclidean dimensions. In three dimensions, the physical properties of this
model are quite different from the four dimensional one. We will show that
many of the interesting features pointed out in \cite{cardy1,cardy2} are also
present in three dimensions. However, there are also significant
differences.
As already explained before, the $\theta$ term becomes important whenever
there are magnetic monopoles or vortex lines. On the lattice we can
naturally define models which contain monopoles and vortex lines as
possible excitations. This is possible if the degrees of freedom are
considered as angular variables. We first present two well known lattice
models which have monopoles and vortex
excitations and then study the effect of
the $\tht$ parameter in these models. We will see that the $\tht$ parameter
couples these two models in a non-trivial way and leads to a rich phase
structure. 
The analysis presented here is based on the technique of duality
transformations as applied to statistical systems. These techniques have
been used very effectively to understand systems like the planar model
\cite{vort} and the $U(1)$ lattice gauge theory \cite{banks}. Recently
in \cite{shar} dual transformations have also been constructed for
non-abelian lattice gauge theories.

The main aim of this paper is to present an analysis of the $\theta$ term in a
lattice model containing both monopoles and vortex lines. The $\theta$ term
is introduced on the lattice by coupling the monopoles to the scalar field
and the vortex lines to the gauge field.
Many of the techniques 
used for studying  the lattice model are well known
in the literature but we
present many details
for the sake of completeness.
Also,
at some places we have managed to give more illuminating derivations of
some of the steps in the analysis.
The organization of this paper is as follows. In Sec.~1 and Sec.~2 we 
discuss lattice models which have monopoles and vortices as possible
excitations.In Sec.~3 we present a detailed analysis of the model obtained
by adding a $\theta$ term. In Sec.~4 we make some concluding remarks.
Some technical details are presented in the appendix.
\end{section}

\begin{section}{A Lattice model with monopoles}
The model considered here is given by the following action
\beq
S_{1}=\frac{-\beta_{g}}{2}\sum_{x\mu \nu}(\partial_{\mu}\phi_{\nu}-
\partial_{\nu} \phi_{\mu}-2\pi\ s_{\mu \nu}(x))^2
+ip\sum_{x\mu}m_{\mu}(x)\phi_{\mu}(x)
\quad .
\eeq
The partition function is given by
\beq
Z_{1}=tr \exp(S_{1})
\quad .
\eeq
The symbol $tr$ denotes
\beq
\sum_{m_{\mu}=-\infty}^{\infty} 
\sum_{s_{\mu \nu}=-\infty}^{\infty} 
\int_{-\infty}^{\infty}d\phi_{\mu}
\quad .
\eeq
The variables $\phi_{\mu}\ \ and\ \ m_{\mu}$
are defined on the links of the lattice and
the integer valued 
variables $s_{\mu \nu}$ are defined on the plaquettes of the lattice.
The symbol $x$ denotes a three-dimensional vector and the symbol
$\partial_{\mu}$ is the lattice derivative.
The fields $\phi_{\mu}$ are the gauge degrees of freedom whereas the
integer valued variables $s_{\mu \nu}$ are the monopole degrees of
freedom. The above model describes gauge fields 
coupled minimally to
current loops with the additional presence of magnetic monopoles.
This model has the following gauge invariance
\beq
\phi_{\mu}(x)\rightarrow \phi_{\mu}+\partial_{\mu}\Lambda
\quad .
\eeq
This also requires that
\beq
\partial_{\mu}m_{\mu}=0
\eeq
Hence the summation over $m_{\mu}$ is restricted to a summation over
closed loops.
To see that this model describes magnetic monopoles consider the
quantity
\beq
F_{\mu \nu}(x)=\partial_{\mu}\phi_{\nu}(x)-\partial_{\nu}\phi_{\mu}(x)-
2\pi s_{\mu \nu}(x)
\quad .
\eeq
Consider a configuration in which $s_{\mu \nu}(x)\ne 0$ on all the plaquettes
pierced by the string in Fig.~\ref{monopole}. This figure shows the projection
of the string on the XY plane (as broken lines) 
and the projection of the plaquettes ( which are in the YZ plane)
on the XY plane (as solid lines).
The string is on the dual
lattice and begins at $P$ and extends to infinity in the $z$ direction.
It is easily seen that for this configuration
\beq
\partial_{\lambda}\tilde F_{\lambda}=-2\pi\delta(x_{P})
\eeq
which means that there is a magnetic monopole at $P$ with a Dirac string
in the $z$ direction. Thus, the integer valued plaquette degrees of freedom
account for magnetic monopoles. The monopole and its associated Dirac
string reside on the dual lattice.
The monopoles in this model can be made explicit by means of a duality
transformation. A duality transformation generally involves three steps.
First, the quadratic part of the action is linearized by introducing an
auxiliary field. The next step is to integrate over the original ( in this
case $\phi_{\mu}$) degrees of freedom and this will lead to a constraint
for $\phi_{\mu}$. This constraint is solved by introducing degrees
of freedom which are the dual variables
and this solution is inserted back into the
partion function. A few manipulations after that lead to the following
form of the partition function
\beqa
\label{z1}
Z_{1}=Tr
\exp \left( -8\pi^{2}\beta_{g}\sum_{x,x^{\prime}}\tilde m(x)G(x-x^{\prime})\tilde
m(x^{\prime})\right)
\exp \left (\frac{-p{2}}{8\beta_{g}}\sum_{x,x^{\prime}}
m_{\mu}(x)G(x-x^{\prime})m_{\mu}(x^{\prime})\right ) \\ \nonumber
X\ 
\exp \left( 2\pi i\sum_{x,x^{\prime}}\tilde m(x)G(x-x^{\prime})\epsilon_{\mu \nu
\lambda}\partial_{\lambda}K_{\mu \nu}^{\star}\right)
\quad .
\eeqa
The monopole density $\mon$ is defined as
\beq
\mon=\frac{1}{2}\epsilon_{\lambda}{\mu}{\nu}\partial_{\lambda}s_{\mu \nu}(x)
\quad .
\eeq
The symbol $\star x$ denotes that the monopole is defined on the dual
lattice point.
This definition requires that the total monopole number in the system is
zero.
The symbol $Tr$ denotes
\beq
\sum_{m_{\mu},\tilde m =-\infty,\infty}
\quad
\eeq
with the understanding that the summation is over configurations
with zero total monopole number and closed loops of currents.
The quantity $K_{\mu \nu}^{\star}$ is a particular solution of
\beq
\partial_{\mu}K_{\mu \nu}^{\star}=p m_{\nu}
\quad .
\eeq
In the above expression for the partition function\  $G(x-x^{\prime})$
is the three-dimensional Green's function
which satisfies
\beq
\label{green}
-\partial^{2}G(x-x^{\prime})=\delta(x-x^{\prime})
\quad .
\eeq
The symbol $\partial^2$ is the lattice laplacian.
The partition function in Eq.~\ref{z1} describes a gas of current loops
($\curr$) and magnetic monopoles ($\mon$). The monopoles and current loops
interact among themselves with a three-dimensional Coulomb interaction.
The last term in Eq.~\ref{z1} describes the interaction a of a monopole
with a current loop. This interaction is just the solid angle subtended 
by the current loop at the monopole and this is shown in the
appendix.
Before proceeding we briefly outline the steps leading to Eq.~\ref{z1}.
The linearization of the action is accomplished by introducing an auxiliary
field $K_{\mu \nu}$ and the partition function becomes
\beq
\label{z1lin}
\int_{-\infty}^{\infty}dK_{\mu \nu} tr
\exp(i\sum_{x\ \mu \nu}K_{\mu \nu}(x)(\partial_{\mu}\phi_{\nu}-
\partial_{\nu}\phi_{\mu}-2\pi\ s_{\mu \nu}(x))
\exp (\frac{-1}{2\beta_{g}}\sum_{x}K_{\mu \nu}^{2}(x))
\exp (ip\sum_{x\mu \nu}m_{\mu}(x)\phi_{\mu}(x))
\quad .
\eeq
Integration over $\phi_{\mu}$ results in the constraint
\beq
2 \partial_{\mu}K_{\mu \nu}=p\ m_{\nu}
\quad .
\eeq
The solution of this constraint is
\beq
\label{constra}
2 K_{\mu \nu}=\epsilon_{\mu \nu \lambda}\partial_{\lambda}\phi
+K_{\mu \nu}^{\star}
\quad .
\eeq
Since $2K_{\mu \nu}$ has to be an integer the fields $\phi(x)$ and $K_{\mu \nu}^
{\star}(x)$ are integer valued. However, the solution to the constraint
is not unique because
\beq
\phi(x) \rightarrow \phi(x)+c
\eeq
also solves the constraint. Using this "dual" gauge invariance
the range of integration over $\phi$ can be taken
from
$-\infty$ to $\infty$. 
Substituting the solution of the constraint in Eq.~\ref{constra} 
in Eq.~\ref{z1lin} and doing a gaussian
integration over $\phi$ leads to Eq.~\ref{z1}.
\end{section}

\begin{section}{A Lattice model with vortices}
A lattice model describing vortices is defined by the action
\beq
S_{2}=-\frac{\beta_{h}}{2}\sum_{x\ \mu}(\partial_{\mu}\theta(x)-
2\pi l_{\mu}(x))^{2}+
+ip\sum_{r}m(r)\theta(r)
\quad .
\eeq
The partition function is given by
\beq
Z_{2}=tr \exp (S_{2})
\quad .
\eeq
The symbol $tr$ denotes
\beq
\sum_{m,l_{\mu}=-\infty}^{\infty}\int_{-\infty}^{\infty}d\theta
\quad .
\eeq
In the above model the variables $\theta$\  and\  $m$
are defined on the sites of the lattice and
the variables $n_{\mu}$ are defined on the links of the lattice.
$\theta$ are the spin-wave degrees of freedom and $n_{\mu}$ are the
vortex degrees of freedom. The above model describes vortex lines interacting
with charged $m(x)$ variables.
The model has the following global invariance
\beq
\theta(x)\rightarrow \theta(x) + c
\eeq
where $c$ is any constant.
This automatically requires
\beq
\sum_{x}m(x)=0
\quad .
\eeq
The vortex lines in this model can be identified by considering the quantity
\beq
V_{\mu}(x)=\partial_{\mu}\theta(x)-2\pi l_{\mu}(x)
\quad .
\eeq
Consider a configuration in which $l_{\mu}$ is non-zero on all the links
pierced by the world line of the string. 
This configuration is also
shown in Fig.~\ref{monopole}.
This configuration is a vortex running in
the $z$ direction which is the direction out of the plane of the figure.
The accompanying string is chosen in the $X$ direction and is indicated by
the broken line which pierces the bonds between pairs of nearest neighbour
sites (shown as solid lines).
Any closed loop about this vortex line will give a
non-zero value for
\beq
\sum_{x}V_{\mu}(x)
\quad .
\eeq
The vortices in this model can again be made explicit by means of a dual
transformation. Introducing an auxiliary field as was done in the previous
section and repeating the procedure described before, we get an expression
for the partition function
\beqa
\label{z2}
Z_{2}=Tr
\exp \left( -\frac{p^{2}}{2 \beta_{h}}\sum_{x,x^{\prime}}m(x)G(x-x^{\prime})
m(x^{\prime})\right)
\exp \left (-2\pi^{2}\beta_{h}\sum_{x,x^{\prime}}
\tilde m_{\mu}(x)G(x-x^{\prime})\tilde m_{\mu}(x^{\prime})\right ) \\ \nonumber
X\ 
\exp \left( -2\pi i\sum_{x,x^{\prime}}\tilde 
m_{\lambda}(x)G(x-x^{\prime})\epsilon_{\lambda \mu
\nu}\partial_{\mu}K_{\nu}^{\star}\right)
\quad .
\eeqa
The $Tr$ denotes
\beq
\sum_{m,\tilde m =-\infty}^{\infty}
\eeq
The vorticity $\vor$ is defined as
\beq
\vor =\epsilon_{\mu \nu \lambda}\partial_{\nu}\tilde \phi_{\lambda}(x)
\quad .
\eeq
From the above equation it is clear that the vortices form closed loops
because
\beq
\partial_{\mu}\vor=0
\quad .
\eeq
As in the case of the monopole, the vortex lines and their associated
strings live on the dual lattice. A closed vortex loop in the
$\mu \nu$ plane will have a sheet swept by its string and the 
plaquettes in this sheet will be dual to the links with $l_{\mu}\ne 0$.
The steps leading to the dual transformation are analogous to those
in the previous section; only the auxiliary field $K_{\mu}$ is now
introduced on every link.
The constraint equation that has to be solved is
\beq
\partial_{\mu}K_{\mu}(x)=pm(x)
\quad .
\eeq
The solution of this constraint equation is
\beq
K_{\mu}(x)=\epsilon_{\mu \nu \lambda}\partial_{\nu}\tilde \phi_{\lambda}
+K_{\mu}^{\star}(x)
\quad ,
\eeq
with $K_{\mu}^\star(x)$ being a solution of the inhomogeneous equation.
The dual gauge invariance in this case is
\beq
\tilde \phi(x)\rightarrow \tilde \phi(x)+\partial_{\lambda}\Lambda
\quad .
\eeq
The first two terms in Eq.~\ref{z2} describe the Coulomb interaction
between $m(x)$ variables and the vortex currents $\vor$. 
($G(x-x^{\prime})$ is the same Green's function as in Eq.~\ref{green})
The last term 
represents an interaction between the vortex currents and the $m$ charges. 
$K_{\mu}^{\star}(x)$ is the solution of the inhomogeneous equation
\beq
\partial_{\mu}K_{\mu}^{\star}(x)=pm(x)
\eeq
This interaction is again proportional to the solid angle 
(apart from a negative sign) of the
vortex current subtended at $m$. A demonstration of this fact is there in
the appendix.

Before we proceed to the model with a $\theta$ term we can already
see that the two models described above have a very similar structure.
For instance, the monopole model has
\beq
\partial_{\mu}m_{\mu}=0
\eeq
as the current conservation equation and
\beq
\sum_{x^\star}\mon=0
\eeq
as the monopole conservation  equation.
In the vortex model, the "m" charge conservation equation is
\beq
\sum_{x}m(x)=0
\eeq
whereas the vortex conservation equation is
\beq
\partial_{\mu}\vor
\quad .
\eeq
The roles of the conservation of ordinary charges and topological
charges are clearly reversed and the two models are dual to each other.
This duality will be made more precise in the next section.
\end{section}
\newpage
\begin{section}{The Coupled Model}
Now we can couple the two previous models by introducing a $\theta$
term as explained in the introduction. The $\theta$ term is defined
by introducing two additional couplings as
\beq
\label{coup}
S_{\theta}=\frac{ip\theta}{2\pi} \sum_{x,\mu}\mon \phi (x)
+\frac{ip\theta}{2\pi} \sum_{x,\mu}\vor \phi_{\mu}(x)
\eeq
The full action of the coupled model is 
given by
\beq
S=S_{1}+S_{2}+S_{\theta}
\quad ,
\eeq
which can be written out as
\beqa
\label{one}
S=\frac{-\beta_{g}}{2}\sum_{x\ \mu \nu}(\partial_{\mu}\phi_{\nu}-
\partial_{\nu}\phi_{\mu}-2\pi s_{\mu \nu}(x))^{2}
-\frac{\beta_{h}}{2}\sum_{x\ \mu}(\partial_{\mu}\theta(x)-
2\pi n_{\mu}(x))^{2}+\\ \nonumber
ip\sum_{x\ \mu}m_{\mu}(x)\phi_{\nu}(r)
+ip\sum_{x}m(x)\theta(x)+
\frac{ip\theta}{2\pi} \sum_{x,\mu}\mon \phi (x)
+\frac{ip\theta}{2\pi} \sum_{x,\mu}\vor \phi_{\mu}(x)
\eeqa
The partition function of this model is given by
\beq
Z=tr \exp (-S)
\quad .
\eeq
The trace represents the following sum over states:
\beq
\sum_{m(x)\ -\infty}^{\infty} \sum_{m_{\mu}(x)\ -\infty}^{\infty}
\int_{-\infty}^{\infty}d\phi_{\mu}(x)
\int_{-\infty}^{\infty}d\theta(x)
\quad .
\eeq
When $\theta=0$ the two models are decoupled and the partition function is
simply a product of their separate partition functions.
\beq
Z=Z_{1}Z_{2}
\quad .
\eeq
The model at $\theta=0$ represents a system of monopoles, currents,
m-charges and vortices. However, the excitations in one system do not interact
with those in the other system.
The model at $\theta=0$ is seen to be trivially self-dual. This follows
by noting that the transformations
\beqa
\label{self}
\beta_{g}\rightarrow \frac{p^2}{16\pi^{2}\beta_{h}} \\ \nonumber
\beta_{h}\rightarrow \frac{p^2}{16\pi^{2}\beta_{g}} \\ \nonumber
m(x) \rightarrow \mon \\ \nonumber
m_{\mu}(x) \rightarrow \vor \\ \nonumber
\eeqa
leave the partition function unchanged.
The dual transformation maps every point on the hyperbola
\beq
\beta_{g} \beta_{h}=\frac{p^2}{16\pi^2}
\eeq
onto itself. The region $\beta_{g} \beta_{h}< \frac{1}{16\pi^2}$
is mapped onto the
region $\beta_{g} \beta_{h}> \frac{1}{16\pi^2}$
by the dual transformation and vice-versa.
However, this self duality property is trivial because the system
on which it acts is a product of two decoupled systems. 
Nonetheless we mention this
fact here because, as we will show later,
at certain values of $\theta$ the self-duality will
still persist
When $\theta \ne 0$
the two systems are coupled in a non-trivial way.
There is a cross coupling between the spin-wave excitaions of one syatem
and the topological excitations of the other system. For instance, the
monopole $\mon$ couples to the spin-wave field $m(x)$ and similarly the
vortex $m_{\mu}(x)$ couples to the gauge field $\phi_{\mu}$.
The main point of this paper is that this coupled system defines an
interesting model that has some exact duality symmetries. It also
has a rich phase structure as a function of $\theta$.
The analysis previously described for the monopole and vortex models
can be repeated in the same way as before. 
The only point to note is that the monopoles and vortices are defined
on the dual lattice whereas the gauge fields and the spin variables are
defined on the original lattice. However, we can approximately take the 
point on the dual
lattice to coincide with the point on the original lattice.
There will be corrections to this approximation but these will involve
higher derivative terms which can only affect the short wavelength
behaviour of the system.
The $\theta$ term changes the
constraint equation for the two auxiliary fields. The new constraint
equations become
\beqa
\partial_{\mu}K_{\mu \nu}=pm_{\nu} +\frac{p\theta}{2\pi}\vor \\ \nonumber
\partial_{\mu}K_{\mu}(x)=pm(x) +\frac{p\theta}{2\pi}\mon
\quad .
\eeqa
The only change is in the inhomogeneous part of these equations
\beqa
\partial_{\mu}K_{\mu \nu}^{\star}=pm_{\nu} +\frac{p\theta}{2\pi}\vor 
\\ \nonumber
\partial_{\mu}K_{\mu}^{\star}(x)=pm(x) +\frac{p\theta}{2\pi}\mon
\quad .
\eeqa
Repeating the steps performed for the monopole or the vortex model
leads to a $\theta$ dependent
partition function
\beqa
\label{ztheta}
Z_{\theta}= tr \\ \nonumber
\exp \left(-\frac{p^{2}}{8\beta_g}\sum_{x x^{\prime}}
(m_{\mu}(x)+\frac{\theta}{2\pi}\tilde m_{\mu}(x)) G(x-x^{\prime})
(m_{\mu}(x^{\prime})+\frac{\theta}{2\pi}\tilde m_{\mu}(x^{\prime})\right )
\\ \nonumber
\exp \left (-\frac{p^{2}}{2\beta_h}\sum_{x x^{\prime}}
(m(x)+\frac{\theta}{2\pi}\tilde m(x)) G(x-x^{\prime})
(m(x^{\prime})+\frac{\theta}{2\pi}\tilde m(x^{\prime}) \right )
\exp \left (-8\pi^{2}\beta_{g}\sum_{x x^{\prime}}\tilde m(x) G(x-x^{\prime})
\tilde m(x^{\prime}\right ) \\ \nonumber
\exp \left (-2\pi^{2}\beta_{h}\sum_{x x^{\prime}}\tilde m_{\mu}(x) 
G(x-x^{\prime})
\tilde m_{\mu}(x^{\prime} \right )
\exp \left( 2\pi i\sum_{x,x^{\prime}}\tilde m(x)G(x-x^{\prime})\epsilon_{\mu \nu
\lambda}\partial_{\lambda}K_{\mu \nu}^{\star}\right)
\\ \nonumber
\exp \left( -2\pi i\sum_{x,x^{\prime}}\tilde 
m_{\lambda}(x)G(x-x^{\prime})\epsilon_{\lambda \mu
\nu}\partial_{\mu}K_{\nu}^{\star}\right)
\quad .
\eeqa
It can be seen that the $\theta$ term couples the gauge and spin models
and introduces additional interactions in each of them.
The partition function of the model is no longer
separable into a spin part and a gauge part and cannot be written as
\beq
Z_{\theta}\ne Z_{1} Z_{2}
\eeq
Hence, the phase structure of this model can be quite complicated.
One of the immediate consequences of the $\theta$ term is that the
term representing the Coulomb interaction between electric loops gets modified.
This means that the vortex loops (for which $\vor \ne 0$) acquire an electric
charge which is given by
\beq
Q_{\mu}(x)=m_{\mu}(x)+\frac{\theta}{2\pi}\vor
\quad .
\eeq
For instance, a loop having only vorticity ($m_{\mu}(x)=0,\ \vor \ne 0$)
will still have an electric charge given by $\frac{\theta}{2\pi}\vor$.
Likewise the Coulomb interaction between $m$ charges is also modified
and acquires a piece due to the monopoles
\beq
Q(x)=m(x)+\frac{\theta}{2\pi}\mon
\quad .
\eeq
Similarly a point having only non-zero monopole density will have an
additional "m" charge $\frac{\theta}{2\pi}\mon$.
The electric charges of the vortex loops and the "m" charges of the
monopoles can now take fractional values because of $\theta$.
It is convenient to associate a vorticity and an electric charge to every
closed loop. The values of these charges are plotted in Fig.~\ref{charglat}.
The first thing to notice about this model is that the partition function
is periodic in $\theta$. This follows from the simple fact that we can always
shift the summed variables $m_{\mu}$ and $m$ as their summation
range is infinite.
In the presence of a $\tht$\  term the model is no longer self-dual
under the transformations in Eq.~\ref{self}. However, for certain
specific values of $\tht$ the model is still self dual. These are the
values for which $2\pi/\theta$ is some integer $q$. 
To see the self duality for these values of $\tht$,
we make a simple change of variables
\begin{eqnarray}
m_{\nu} + \frac{\tilde m_{\nu}}{q} =-\frac{\tilde l_{\nu}}{q} \\ \nonumber
m + \frac{\tilde m}{q} =-\frac{\tilde l}{q}
\quad .
\end{eqnarray}
Expressing the partition function in terms of $\tilde l_{\nu}$,$m_{\nu}$,
$\tilde l$ and $m$, the partition function reduces to the
the original one provided we identify $\tilde l_{\mu}$
with $\tilde m_{\mu}$, $\tilde l$ with $\tilde m$,
and make the following changes.
\begin{eqnarray*}
\label{newf}
\beta_{g} \rightarrow  \frac{p^2}{16\pi^2 q^{2}\beta_h} \\ \nonumber
\beta_{h} \rightarrow  \frac{p^2}{16\pi^2 q^{2}\beta_g}
\quad .
\end{eqnarray*}
Unlike the dual transformations in Eq.~\ref{self} the above dual
transformations contain more information because they now act on
a system which is not a decoupled system. These dual transformations
can be used to understand the phase diagram of the model in one region
if the phase diagram is known in another region. The points of the
phase diagram which are left invariant under the dual transformation are
the points on the hyperbola
\beq
\beta_{g} \beta_{h} =\frac{p^2}{16\pi^2 q^2}
\quad .
\eeq
The self-duality in Eq.~\ref{newf} holds true only when
$\theta=
2\pi/q$,$q$ being an integer. We now point out another symmetry that
is present in the model for arbitrary values of $\theta$.
This symmetry can be deduced by requiring the partition
function to be invariant under the
following transformations:
\begin{eqnarray}
m \rightarrow -\tilde m \\ \nonumber
m_{\mu} \rightarrow -\tilde m_{\mu} \\ \nonumber
\beta_{g} \rightarrow \beta_{g}^{\prime} \\ \nonumber
\beta_{h} \rightarrow \beta_{h}^{\prime} \\ \nonumber
\theta \rightarrow \theta^{\prime} \\ \nonumber
\quad .
\end{eqnarray}
As the partition function is a trace over the $m,\tilde m,m_{\mu},
\tilde m_{\mu}$ degrees of freedom it can be written as
\beq
Z
=Z_{\theta}({m,\tilde m,m_{\mu},\tilde m_{\mu}})
\eeq
Imposing the condition
\beq
Z_{\theta^\prime}({-\tilde m,- m,\tilde m_{\mu}, m_{\nu}})
=Z_{\theta}({m,\tilde m,m_{\mu},\tilde m_{\mu}})
\eeq
we get the following set of equations
\begin{eqnarray}
8\pi^2 \beta_{g}+\frac{p^2 \theta^2}{8\pi^2 \beta_{h}}=
\frac{p^2}{2\beta_{h}^{\prime}} \\ \nonumber
\frac{p^2}{2\beta_{h}}=8\pi^2 \beta_{g}^{\prime}+\frac{p^2 \theta^{\prime}}
{8\pi^2 \beta_{h}^{\prime}} \\ \nonumber
2\pi^2 \beta_{h}+\frac{p^2 \theta^2}{32 \pi^2 \beta_{g}}=\frac{p^2}
{8\beta_{g}^{\prime}} \\ \nonumber
\frac{p^2}{8\beta_{g}}=\frac{p^2 {\theta^{\prime}}^2}{32 
\pi^2 \beta_{g}^{\prime}}
+2\pi^2 \beta_{h}^{\prime}
\end{eqnarray}
We get four equations in the three variables, $\beta_{g},\beta_{h}$, and
$\theta$. Interestingly, these four equations are consistent and have the
following solution:
\begin{eqnarray}
\label{dueqns}
\beta_{h}^{\prime}=\frac{4\pi^2\beta_{h}p^2}{64\pi^4\beta_{g}\beta_{h}+
p^2 \theta^2} \\ \nonumber
\beta_{g}^{\prime}=\frac{4\pi^2\beta_{g}p^2}{64\pi^4\beta_{g}\beta_{h}+
p^2 \theta^2} \\ \nonumber
{\theta^{\prime}}^2=\frac{16\pi^4\theta^2 p^4}{(64\pi^4\beta_{g} \beta_{h}
+p^2 \theta^2)^2}
\end{eqnarray}
This means that we can define an exact duality symmetry in the model which
now acts on a three coupling space.
The duality equations can be recast in a more concise form by defining
the variable
\beq
\label{newvar}
z=\sqrt(\beta_{g} \beta_{h})
\quad .
\eeq
In terms of the variables $z$ and $\theta$, the duality equations are
\begin{eqnarray}
{z^{\prime}}^2=\frac{16\pi^{4}p^{4}z^2}{(64\pi^4 z^2 +p^2 \theta^2)^2} 
\\ \nonumber
{\theta^{\prime}}^2=\frac{16 \pi^{4} p^4 \theta^2}{(64\pi^4 z^2 
+p^2 \theta^2)^2}
\quad .
\end{eqnarray}
If we define the complex coupling $\zeta$ by
\beq
\zeta=\frac{4\pi z}{p}+\frac{i\theta}{2\pi}
\quad ,
\eeq
the duality transformation can be expressed as
\beq
\zeta \rightarrow \frac{1}{\zeta}
\quad .
\eeq
Since the action is also periodic in $\theta$ with period $2\pi$,
the transformation
\beq
\zeta \rightarrow \zeta + i
\eeq
is also a symmetry of the model.
These two transformations do not commute with each other and generate
the group $SL(2,Z)$. This symmetry group was first pointed out in
\cite{cardy2} in the four dimensional model. In the three-dimensional
model that we are considering here, the duality symmetry has a slightly
more complicated form as in
Eq.~\ref{dueqns} and the group $SL(2,Z)$ acts on the variable
$z$ which is given by Eq.~\ref{newvar}.

We now proceed to study the phase structure of this model.
The phase structure of this model will be a function of the three
parameters $\beta_{g}$,$\beta_{h}$ and $\theta$ at a given value of $p$.
The various phases of the model will be characterized by the behaviour of
the the $m(x),\tilde m(x),m_{\mu}(x),\tilde m_{\mu}(x)$ excitations.
Depending on the density of these excitations, the phases in
this model will have different physical properties.
In order to arrive at the phase structure of the model we will use
simple free energy arguments based on the energy and the entropy of loops.
Though these arguments are admittedly crude and ought to be substantiated
by other methods they provide us with a qualitative picture of the phase
structure.
We first note that there will be regions
in parameter space where the loops will condense. There are two kinds
of loops that can condense. They are labelled by $m_{\mu}(r)$ and
$\tilde m_{\mu}(r)$;
$m_{\mu}(r)$ loops are
the world lines of electrically charged particles,$\tilde m_{\mu}(r)$ loops
are the vortex excitations.
A composite loop having both $m_{\mu}(r)$ and $\tilde m_{\mu}(r)$ is
also possible and it will be referred to as a composite loop.
These composite loops are formed by the binding of a vortex to an
elecrtric current loop.
In addition to these loops there are also the point like excitations
labelled by
$m(x)$ and $\tilde m(x)$. $\tilde m(x)$s are the
magnetic monopoles and
$m(x)$s will be referred to as spin charges.
The reason for this terminology is that the
$m(x)$s represent source terms for the spin variables.
In addition to these excitations there are also excitations which
simultaneously carry magnetic charge and a spin charge. These composite
objects will be referred to as composite charges. 
The composite charges are formed by the binding of a magnetic monopole
to a spin charge.
A composite loop or a composite charge is a combination of an ordinary
charge and a dual excitation.
If we neglect the long range Coulomb interaction, a crude
estimate for the free energy of a loop 
of length $L$ having charges $(m_{\mu},\tilde m_{\mu})$ is
\beq
\label{fengy}
F(L)=(\frac{2\pi^2 p^2}{\beta_g}(m+\frac{\theta}{2\pi}\tilde m)^2
+ 2\pi^2 \beta_{h}(\tilde m)^2)G(0)L -(\log c) L
\quad .
\eeq
In the above approximation, only the self interaction of the loop
contributes to its energy.
Condensation of loops having charges $(m_{\mu},\tilde m_{\mu})$
occurs whenever the loop free energy becomes negative.
The constant $c$ depends on the co-ordination number of the lattice and is
approximately $5$ for a three-dimensional cubic lattice.
In the above approximation only the self energy of the loop has been
considered in the expression for the free energy. Though this is quite a
severe approximation, we expect it to reproduce
the general features of the phase diagram.
As we have already noted,
the effect of the $\theta$ term is to give an electric charge to
the vortex lines as
\beq
\label{dyloop}
q_{\nu}=m_{\nu} +\frac{\theta}{2\pi} \tilde m_{\nu}
\quad .
\eeq
Similarly, the magnetic monopoles get an effective spin charge
\beq
\label{dyp}
q=m +\frac{\theta}{2\pi} \tilde m
\quad .
\eeq
We first consider the phase diagram at $\theta=0$. When $\theta=0$ the
partition function can be decoupled into two partition functions
\beq
Z=Z_{1}Z_{2}
\eeq
and each of these systems can be considered individually.
If we consider $Z_{1}$ first, the free energy of a
loop of length $L$ carrying
electric charge $m_{\mu}$ is given by
\beq
F(L)=\frac{2\pi^2 p^2}{\beta_{g}}m^{2} G(0)L -(\log c) L
\quad .
\eeq
Thus $F(L)<0$ if
\beq
\beta_{g}>\frac{2\pi^2 p^2 G(0)}{\log c}
\quad .
\eeq 
Therefore the model described by $Z_{1}$ will exist in two phases,
a small $\beta_{g}$ phase in which the current loops are very sparse
and a large $\beta_{g}$ phase in which the current loops are very dense.
The other excitations present in this model, namely the magnetic monopoles
will have a density given by
\beq
\rho(\beta_{g})=\exp(-8 \pi^2 \beta_{g}G(0)\tilde m^{2}(x))
\quad .
\eeq
Unlike the density of current loops, the magnetic monopole density
falls continuously to zero as the coupling constant $\beta_{g}$ is increased.
We can make a similar analysis of the model $Z_{2}$. The same free energy
arguments applied to the vortex loops gives $F(L)<0$ if
\beq
\beta_{h}<\frac{\log(c)}{2\pi^2 G(0)}
\quad .
\eeq
The small $\beta_{h}$ phase has a high density of vortex loops whereas
the large $\beta_{h}$ phase has a very low density of vortex loops.
The spin charges $m$ change continuously with $\beta_{h}$ as
\beq
\rho(\beta_{h})=\exp(-\frac{p^2 G(0)m^2}{ 2\beta_{h}})
\quad .
\eeq
For the coupled model the analysis can be repeated but there are now
three coupling constants $\beta_{g}$,$\beta_{h}$ and $\theta$. 
We consider the condensation condition given by Eq.~\ref{fengy} for
various limiting values of $\beta_{g}$ and $\beta_{h}$.
It we fix $\theta=\pi$ the different excitations present in the system
are those which correspond to the black points in the charge
lattice in Fig.~\ref{charglat}. 
The excitations corresponding to these black points are
multiples of these fundamental excitations.

\noindent
1.Electric current loops $(1,0)$:\ These have a non-zero electric charge. 
\\
2.Vortex loops $(0,1)$:\ These have a non-zero electric charge because of
Eq.~\ref{dyloop}.
\\
3.Composite loops $(1,-2)$:\ These loops have exactly zero 
electric charge 
again because of Eq.~\ref{dyloop}.
\\

\noindent
A similar charge lattice can be drawn for the point like excitations
in the model. The excitations of this charge lattice are multiples
of the following fundamental charges.

\noindent
1.Spin charges $(1,0)$:\ These have only "m" charge.\\
2.Magnetic monopoles $(0,1)$:\ These have a non-zero "m" charge because
of Eq.~\ref{dyp}.\\
3.Compositely charged point like objects $(1,-2)$:\ These have a zero
"m" charge because of Eq.~\ref{dyp}.

We have to compare the free energies of these condensates and then choose
the one with the lowest free energy.
First we can explore various limits of the coupling space.

\noindent
1.$\beta_{h}=0$. \\
The free energies of the possible condensates are given by
\beq
F(L)/L=\frac{p^2}{8\beta_g}G(0)(m+\frac{1}{2}\tilde m)^2 -\log (c)
\quad .
\eeq
(a) $(1,0) :\frac{p^2}{8\beta_{g}}G(0) -\log(c)$
\\ \nonumber
(b) $(0,1): \frac{1}{32\beta_{g}}G(0) -\log(c)$
\\ \nonumber
(c) $(1,-2): -\log(c)$

\noindent
The above estimates show that the composite loops $(1,-2)$ will always have
the lowest free energy. Therefore, in this limit, there is always a condensate
of composite loops $(1,-2)$ and there will be no phase
transition on this axis.

\noindent
2.$\beta_{h}=\infty$
The general free energy relation Eq.~\ref{fengy} shows that this limit
forces $\tilde m_{\mu}\approx 0$. The free energy condition becomes
\beq
F(L)=\frac{p^2}{8\beta_{g}}m^2 G(0)-\log(c)
\quad .
\eeq
Since $\tilde m$ is forced to be zero we only have to consider
electric
$(1,0)$ loops. $F(L)$ of $(1,0)$ loops becomes negative
for $\beta_{g}>\frac{G(0)}{8\log(c)}$.
This implies that there is a phase transition on this axis from a small
$\beta_{g}$ phase containing very few current loops to a large
$\beta_{g}$ phase containing very large current loops.

\noindent
3.$\beta_{g}=0$.
In this limit we get the following constraint
\beq
(m+\frac{1}{2}\tilde m)=0
\quad .
\eeq
This leaves only the composite loops $(1,-2)$ as possible excitations.
The free energy condition becomes
\beq
F(L)=8\pi^2\beta_{h}G(0) -\log(c)
\quad .
\eeq
Thus, $(1,-2)$ loops will condense for
\beq
\beta_{h}<\frac{\log(c)}{8\pi^2 G(0)}
\quad .
\eeq
Therefore, on this axis we expect a small $\beta_{h}$ phase
in which $(1,-2)$ loops condense and a large $\beta_{h}$ phase in which
the density of $(1,-2)$ loops is very small.

\noindent
4.$\beta_{g}=\infty$.
The free energy condition is
\beq
F(L)=2\pi^2 \beta_{h}{\tilde m}^2 G(0) -\log(c)
\quad .
\eeq
(a)$(1,0):
F(L)=-\log(c)$ \\
\noindent
(b) $(0,1):
F(L)=2\pi^2 \beta_{h} G(0) -\log(c)$ \\
\noindent
(c) $(1,-2):
F(L)=8\pi^2 \beta_{h}G(0) -\log(c)$ \\
\noindent
It is clear that the condensate $(1,0)$ always has the lowest free energy.
Therefore, the electric current loops are always dense on this axis and there is
no phase transition on this axis.
The other special case to be considered is \\ \noindent
5.$\beta_{g}=\beta_{h}$ \\ \noindent
In this limit the free energy condition becomes
\beq
F(L)=\frac{p^2}{8\beta}(m+\frac{1}{2}\tilde m)^2 G(0)
+2\pi^2 \beta {\tilde m}^2 G(0) - \log(c)
\quad .
\eeq
\end{section}
The condensation criterion can be written as the interior of the
ellipse on the charge lattice
\beq
(\frac{q^2}{a^2}+\frac{{\tilde m}^2}{b^2})<1
\quad .
\eeq
The major and minor axes of the ellipse are given by
\beqa
a^2=\frac{8\beta \log(c)}{G(0)p^2} \\ \nonumber
b^2=\frac{\log(c)}{2\pi^2 \beta G(0)} \\ \nonumber
\eeqa
When $a>>1$ and $b<<1$ the ellipse is very flat along the q-axis and the
condensate with the lowest free energy is $(1,0)$. When $a<<1$ and $b>>1$
the ellipse is very flat along the $\tilde m$ axis and the condensate with
the lowest free energy is $(1,-2)$. For values of $a$ and $b$ such that
they are comparable, the free energy is the lowest for the $(0,1)$ condensate.
Therefore we generically expect three phase transitions on this axis.
This case occurs in \cite{cardy1} in the four-dimensional
model. It is interesting to note that exactly the same 
condensation condition appears
in the three-dimensional model with the
only difference being that the four-dimensional Green's function gets
replaced by the three-dimensional one and the entropy factor takes
different value.

Now that we have discussed the condensates in the various regions
of parameter space,
we can propose the phase diagram of this model
at $\theta=\pi$. It is given in Fig.~\ref{phase}. So far we have only
discussed phase transitions of the electric, vortex or composite loops.
We have already noted that there are other excitations present in this
model. These are the point like excitations namely the magnetic monopoles,
"m" charges and the composite charges of $m$ and $\tilde m$.The density of these
excitations changes continuously with $\beta_{g},\beta_{h}$ and $\theta$.
The density of the magnetic monopoles and the "m" charges are given
by
\beqa
\label{pointex}
\rho_{m,\tilde m}=\exp (-8\pi^2 \beta_{g}G(0){\tilde m}^2)
\exp(\frac{-p^2}{2\beta_{h}}G(0)(m+\frac{1}{2}\tilde m)^2)
\quad .
\eeqa
This is one of the special features of the model in three-dimensions.
As these excitations are point like, they form a three-dimensional
Coulomb gas of
charged objects whose density is always non-zero. 
The densities of the three possible types of point like excitations
in this model are give by \\
1.$(0,1)$: Magnetic monopoles
\beq
\rho_{\tilde m}\approx \exp -(8\pi^2 \beta_{g}+\frac{p^2}{8\beta_{h}})G(0)
\quad .
\eeq
2.$(1,0)$: Spin charges
\beq
\rho_{m}\approx \exp (-\frac{p^2}{2\beta_{h}}G(0))
\eeq
3.$(1,-2)$: Composite charges
\beq
\rho_{c}\approx \exp (-64\pi^2 \beta_{g} G(0))
\eeq
Since each of these densities is always non-zero, we have a three-
dimensional Coulomb gas containing three species of charged particles.
The densities of these
three species of charged particles determine various
correlation functions (which are defined later on) independently
and can lead to different  effects at different values in the
parameter space.
The behaviour of physical correlation functions 
will be determined by
the density of electric and vortex loops as well as the density of
magnetic monopoles
and "m" charges.
The various regions in the schematic phase diagram in Fig.~\ref{phase}
are discussed
below. \\ 
\noindent
A. In this region only the composite loops $(1,-2)$ condense.
At large values of $\beta_{h}$, the density of magnetic monopoles and
spin charges is large small whereas at small values of $\beta_{h}$
they become non-zero but obey the relation
\beq
m+\frac{\tilde m}{2}=0
\quad .
\eeq
The density of magnetic monopoles decreases as $\beta_{g}$ is increased.

\noindent
B. In this region the electric current loops $(1,0)$, the vortex
loops $(1,0)$, and the composite loops $(1,-2)$ 
have almost zero density. Therefore, this phase is essentially
free of all loop excitations.
The magnetic monopoles $(0,1)$
and spin charges $(1,0)$ have a large density at the top left corner
of the phase diagram which decreases as
as we go away from it in any direction. 

\noindent
C. In this region the electric loops $(1,0)$ condense whereas vortex loops
$(0,1)$ have very low density. The magnetic monopoles $(0,1)$
have a very low density and the spin charges $(1,0)$ have a large
density at the top tight hand corner of the phase diagram.

\noindent
D.In this region the vortex loops $(0,1)$ condense and 
the electric current loops
$(1,0)$ have a very low density. The magnetic monopoles $(0,1)$ and
spin charges $(1,0)$ both have a density which is not very large or very
small. 

\noindent

We will now show that these different phases can be characterized by
the behaviour of
correlation functions which are of the order disorder type.
Before we do this we briefly describe how these
correlation functions are defined in each of the models discussed in the
previous two sections. The correlation functions are simple generalizations
of the Wilson loop \cite{wils}.
In the model described in Eq.~\ref{z1} we can introduce an external
current $J_{\mu}(x)$ in a loop $C$ on the lattice
and a monopole-antimonopole pair at points
$\star x_{1}$ and $\star x_{2}$. The monopole pair is introduced by choosing
$t_{\mu \nu}=1$ on a string joining $\star x_{1}$ and 
$\star x_{2}$. The
correlation function is defined by
\beq
W(C,x_{1},x_{2})=\frac{Z_{1}(m_{\mu}\rightarrow m_{\mu}+J_{\mu},
s_{\mu \nu}\rightarrow s_{\mu \nu}+s_{\mu \nu}^{\prime}(x))}
{Z_{1}}
\quad .
\eeq
This is equivalent to making the following change in $\mon$
\beq
\mon\rightarrow \mon+\rho(\star x)
\eeq
where $\rho$ is defined as
\beq
\rho(\star x)=\delta(\star x-\star x_{1})-\delta(\star x-\star x_{2})
\quad .
\eeq
The correlation function $W$ measures the free energy of an external
current loop  $J_{\mu}$ on $C$ and an external monopole-antimonopole pair
at points $\star x_{1}$ and $\star x_{2}$.
We can similarly define a correlation function for the model described in
Eq.~\ref{z2} as
\beq
\tilde W(\star C,x_{1},x_{2})=\frac{Z_{2}(m\rightarrow m+M(x),l_{\mu}\rightarrow
l_{\mu}(x)+l_{\mu}^{\prime}(x))}{Z_{2}}
\quad .
\eeq
Here $l_{\mu}^{\prime}\ne 0$ on all the links which are dual to the
plaquettes in a surface bounded by
the loop $\star C$ on the dual lattice and $M(x)$ has
the form
\beq
M(x)=\delta(x-x_{1})-\delta(x-x_{2})
\quad .
\eeq
The correlation function $\tilde W$ measures the free energy of
an external vortex loop $\star C$ and an external pair of spin charges at $x_{1}$
and $x_{2}$.
The correlation functions $W$ and $\tilde W$ are examples of order
disorder variables as they involve both the gauge(spin) degrees of freedom
and the monopole(vortex) degrees of freedom.
In the interacting model we can consider the following correlation function
\beq
W_{int}(C,\star C,x_{1},x_{2},\star x_{1},\star x_{2})=
\frac{Z(m_{\mu} \rightarrow m_{\mu} +J_{\mu},s_{\mu \nu} \rightarrow s_{\mu \nu}
+s_{\mu \nu}^{\prime},m\rightarrow m+M,l_{\mu} \rightarrow l_{\mu}+
l_{\mu}^\prime)}{Z}
\quad .
\eeq
In the presence of interactions (when $\theta\ne 0$)
\beq
W_{in}\ne W \tilde W
\eeq
where we have suppressed the arguments of the correlation function.
This correlation function obeys the same duality invariance as the
partition function
\beq
W_{int}(C,\star C,x_{1},x_{2},\star x_{1},\star x_{2})_{\beta_{g},\beta_{h},
\theta}=
W_{int}(\star C,C,\star x_{1},\star x_{2},x_{1},x_{2})_{\beta_{g}^\prime,
\beta_{h}^{\prime},\theta^{\prime}}
\eeq
where $\beta_{g}^\prime$,$\beta_{h}^\prime$ and $\theta^{\prime}$ are related
to the unprimed values by the duality equations in Eq.~\ref{dueqns}.

It is well known that a dilute gas of magnetic monopoles results in an area
law behaviour for the Wilson loop \cite{poly}. This effect arises from
the form of the monopole-current loop interaction. 
Also, an external monopole-antimonopole pair will experience a screened
Coulomb potential in this gas.
Since the form of the
spin charge-vortex loop interaction is the same as the monopole-current loop
interaction, a dilute gas of spin charges will result in a similar
area law behaviour for an external vortex loop. 
Again, an external pair of m charges will experience a screened
Coulomb interaction in this gas.
Similarly, composite charges $(1,-2)$
will result in an area law for an external composite Wilson loop that consists of
a loop of electric charge (1) and vorticity (-2).
However our model also has loop like excitations which can screen
external Wilson loops and vorticity loops which can screen external
vortex loops.
For instance, a condensate of 
$(1,0)$ loops can screen a Wilson loop of charge $(1,0)$ 
and yield a perimeter law. When
both current loops and monopoles are present, the Wilson loop will have an
area law piece but the perimeter law piece will dominate at large
distances. Similarly,
when $(0,1)$ vortex loops condense an external vortex loop of charge $(0,1)$
can be screened by these loops and again result in a perimeter law for
the external vortex loop inspite of the the presence of $m$ charges which
alone would have yielded an area law. The same holds true for
composite $(1,-2)$ loops and $(1,-2)$ charges.
From the partition function in Eq.~\ref{ztheta} we see that there is no
monopole-vortex  or spin charge-current loop interaction. Thus these
excitations cannot influence each other in any drastic
way.

\noindent

\begin{section}{Conclusions}
In this paper we showed that it is possible to define an  analogue of the
$\theta$ term in
three dimensions. We presented our analysis of the $\theta$ term in a 
lattice model which was
motivated by the work of the authors in
\cite{cardy1,cardy2}.
The $\theta$ term couples magnetic monopoles to the scalar
field and vortices to the gauge field. 
The phase structure of the model
changes as a function of $\theta$. In fact, the interactions in this
model arise entirely because of the non-zero value of $\theta$. This model
has an exact duality symmetry which acts on a three parameter space. This
seems to be the first example of a statistical model
in which the duality transformation acts on a three coupling space. 
We also made a qualitative analysis of the phase diagram using energy
entropy arguments and showed that at non-zero values of $\theta$ there
are phases in which excitations having composite loops condense.
A special feature of the three-dimensional model is that there are
novel point like excitations of different species which form a three
dimensional Coulomb gas.
Our analysis is also instructive in understanding how the dual
transformation works in systems containing both vortices and monopoles.
Since the phase diagram of this model (which was studied at $\theta=\pi$)
admits phases with charged vortices, it will be interesting if some
condensed matter systems can be described by this model.
Another interesting avenue is to introduce $\theta$ like terms in non-abelian
gauge theories and look for oblique phases at non-zero values of $\theta$.
	
\end{section}
\vspace{0.5cm}
\newpage
\section{Appendix A}
In this appendix we will examine the form of the monopole-
current loop and the vortex-m charge interaction.
The monopole-current loop interaction is given by
\beq
\label{int1}
\sum_{x,x^{\prime}}\mon G(x-x^{\prime})\epsilon_{\mu \nu 
\lambda}\partial_{\mu}K_{\nu \lambda}(x^{\prime})
\eeq
$K_{\mu \nu}^{\star}$ is a particular solution of
\beq
\label{inh}
\partial_{\mu} K_{\mu \nu}^{\star}=pm_{\nu}
\eeq
Consider a configuration in which the monopole is along the $z-axis$ and
at a distance $z$ and there is a circular 
current loop (of radius R) is in the X-Y plane i.e only
$m_{x},m_{y}\ne 0$. It is easily seen that a particular solution of
Eq.~\ref{inh} is
\beqa
K_{xy}=-p\ \quad inside C \\ \nonumber
K_{xy}=0\ \quad otherwise \\ \nonumber
K_{xz}=K{yz}=0 \\ \nonumber
\quad .
\eeqa
This reduces the monopole-current loop interaction to
\beq
-2\sum_{x,x^{\prime}}\mon \partial_{z^{\prime}}G(x-x^{\prime})K_{xy}^{\star}(x^
\prime)
\eeq
The co-ordinates of $x$ and $x^{\prime}$ are given by
\beqa
x\ \ (0,0,z) \\ \nonumber
x^{\prime}\ \ (x^{\prime},y^{\prime},0)
\eeqa
Using 
\beq
\label{gree}
\partial_{i^{\prime}}G(x-x^{\prime})=\frac{(x-x^{\prime})_{i}}{|x-x^{\prime}|^3}
\eeq
the interaction is given by the expression
\beq
2pm\int_{s}\frac{z\ dx^{\prime} dy^{\prime}}
{({x^{\prime}}^2+{y^{\prime}}^2+{z^\prime}^2)^{3/2}}
\eeq
The integration is over the area of the loop.
The evaluation of this integral is straightforward and gives
\beq
I=2pm(2\pi (1-\cos(\alpha))
\eeq
where $\alpha$ is the azimuthal subtended by the current loop at the monopole.

Now we turn to the vortex-m interaction. It has the form
\beq
-\sum_{x,x^{\prime}} \tilde m_{\lambda}(x) G(x-x^{\prime})\epsilon_{\lambda \mu
\nu}\partial_{\mu}K_{\nu}^{\star}(x^{\prime})
\eeq
$K_{\mu}^{\star}$ is the solution of the inhomogeneous equation
\beq
\partial_{\mu}K_{\mu}^{\star}(x)=pm(x)
\eeq
Consider a configuration in which $m(x)\ne 0$ at a point on the $z$ axis a
distance $L$ from the origin.
The solution of the inhomogeneous equation is chosen to be
\beqa
K_{x}^{\star}=K_{y}^{\star}=0 \\ \nonumber
K_{z}^{\star}=p\Theta(z-L)
\quad .
\eeqa
The $\Theta$ function has the property
\beqa
\Theta(z)=1 \quad z>0 \\ \nonumber
\Theta(z)=0 \quad z<0
\quad .
\eeqa
The co-ordinates of the points $x$ and $x^{\prime}$ are given by
\beqa
x\quad (R\cos(\theta),R\sin(\theta),0) \\ \nonumber
x^{\prime} \quad (0,0,z^{\prime})
\eeqa
Again using Eq.~\ref{gree} we are led to the integral
\beq
-2\pi mp R^{2} \int_{L}^{\infty}\frac{dz^{\prime}}{(R^{2}+{z^{\prime}}^2)^{3/2}}
\eeq
This integral is also straightforward to evaluate and gives the solid angle
interaction
\beq
I=-2pm(2\pi(1-Cos(\alpha))
\eeq
which is just the negative of the monopole-current loop interaction.

\noindent The purpose of this appendix was to show that the monopole-
current loop interaction has the same form as the interaction between
the $m$ charge and the vortex loop apart from a sign factor.

\newpage
\begin{figure}
\label{monopole}
\caption{The string attached to the dual excitations runs on the
dual lattice.}
\end{figure}
\begin{figure}
\label{charglat}
\caption{Charge lattice at $\theta=\pi$.}
\end{figure}
\begin{figure}
\label{phase}
\caption{Schematic phase diagram at $\theta=\pi$.The X-axis is the $\beta_{g}$
coupling and the Y-axis is the $\beta_{h}$ coupling.}
\end{figure}

\newpage
\begin{thebibliography}{99}
\bibitem{vort}{J.~Jose, L.~Kadanoff,S.~Kirkpatrick, and 
D.~Nelson,Phys. Rev.{\bf
B 16}, 1217 (1977).}
\bibitem{hopol}{G.~'t Hooft, Nucl.Phys.{\bf B79}, 276 (1974); A.~M~Polyakov,
 JETP. Lett. \bf{20}, 194 (1974).}
\bibitem{poly}{A.~M.~Polyakov, Nucl.Phys.{\bf 120},429 (1977);
A.~M.~Polyakov, Phys. Lett. {\bf B59},82 (1975).}
\bibitem{instant}{A.~A.~Belavin, A.~M.~Polyakov,A.~S.~Shvartz, and
Yu.~S.~Tyupkin,
Phys. Lett. {\bf B59}, 85 (1974).}
\bibitem{witt}{E.~Witten, Phys. Lett.{\bf B86},283 (1978).}
\bibitem{conj}{G.~t Hooft, Nucl. Phys. {\bf B190}, 455 (1981).}
\bibitem{cardy1}{J.~Cardy and E.~Rabinovici, Nucl. Phys. {\bf B205},1 (1982).}
\bibitem{cardy2}{J.~Cardy, Nucl. Phys. {\bf B205},17 (1982).}
\bibitem{sav}{R.~Savit, Rev. Mod. Phys. {\bf 41}, 1 (1978). }
\bibitem{banks}{T.~Banks, R.~Myerson, and J.~Kogut, Nucl. Phys. {\bf B129},
493 (1977).}
\bibitem{shar}{R.~Anishetty, S.~Cheluvaraja, M.~Mathur, and
H.~Sharatchandra,
Phys. Lett. {\bf B314}, 387 (1993).}
\bibitem{wils}{K.~G.~Wilson, Phys. Rev. {\bf D10},2445 (1974).}
\end {thebibliography}
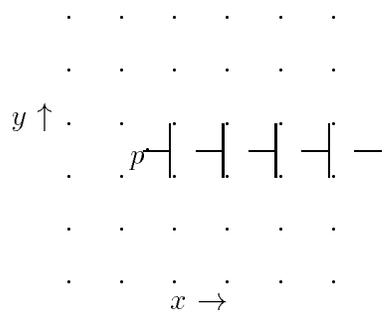
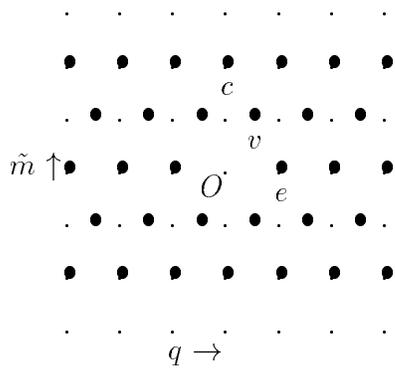
\end{document}